\documentclass{PoS}

\title{The supernovae associated with gamma-ray bursts}

\ShortTitle{SNe of GRBs}

\author{\speaker{David Bersier}\\  
        Astrophysics Research Institute\\
        Liverpool John Moores University\\
        E-mail: \email{dfb@astro.livjm.ac.uk}}

\abstract{The connection between long GRBs and supernovae is now well
  established. I briefly review the evidence in favor of this
  connection and summarise where we are observationally.  I also use a
  few events to exemplify what should be done and what type of data
  are needed.  I also look at what we can learn from looking at SNe
  not associated with GRBs and see how GRBs fit into the broad picture
  of stellar explosions.

}

\FullConference{Gamma-Ray Bursts 2012 Conference -GRB2012,\\
		May 07-11, 2012\\
		Munich, Germany}

\begin{document}

\section{Introduction}

The connection between gamma-ray bursts (GRBs) and supernovae (SNe)
has existed for longer than the ``official'' history of GRBs.  Colgate
\cite{Colgate68} predicted that exploding stars should emit $\gamma$
rays. Even though it turned out that this model for GRB prompt
emission is not correct, it meant that a connection between SNe and
GRBs was on everybody's mind from the beginning. Indeed, Klebesadel,
Strong \& Olson \cite{KSO73} searched for spatial and temporal
coincidence between explosive events (novae and supernovae) and
GRBs. Although they didn't find any coinciding event, this was not
enough to discourage further similar searches.  It is only in the
afterglow era that major progress could finally be made and that a
clear link between long GRBs and SNe would be firmly established.

This review is not an attempt to be exhaustive so there is no
discussion of all known or candidate SNe associated with GRB. I will
rather try to show where we are now as a community and to place these
SNe in a general context. Several recent reviews took a close look at
the evidence for each SN/GRB and should be consulted for specific
information about any particular object (see e.g. \cite{WB06, HB12}).

\section{From suspicion to proof}

GRB980425 was quickly associated with a supernova, SN~1998bw. The
spectra of the SN clearly showed that it was a type Ic -- meaning that
the spectrum showed no sign of hydrogen nor helium. The association
between the GRB and the SN meant that the burst was extremely faint
however ($E_{iso} < 10^{48}$ erg).  SN~1998bw itself was also
peculiar, as it was a very bright SN, showed very broad lines (BL) and
had a large expansion velocity (e.g. \cite{Galama98}.  Because of this
highly unusual nature -- very weak and very nearby while the SN was
very bright, there was still a question mark over this event and some
fraction of the community was still waiting for a GRB at a genuinely
``cosmological'' distance.

This happened with GRB030329. The burst was close enough that a SN
could be seen, yet it was distant and bright enough that it could be
classified as a genuine GRB. Unambiguous SN features became visible in
afterglow spectra a few days after the burst (\cite{Stanek03,
  Hjorth03}). The SN was shown to be very similar to SN~1998bw
\cite{Matheson03} and was also classified as a SN~Ic-BL.  This event
removed any doubt as to the association between SNe and GRBs. 

Since then, several other GRBs have had indisputable spectral evidence
supporting the presence of a SN (see Table~\ref{tab:sne}). Several
good quality spectra (between 2 and many) have been obtained also for
GRB031203 (SN~2003lw \cite{Malesani04}), XRF060218 (SN~2006aj
\cite{Pian06}), GRB100316D (SN~2010bh, \cite{Chornock10, Bufano12}),
GRB120422A (SN~2012bz, \cite{Malesani12, Sanchez12})\footnote{This
  last event is still a developing story as of this writing. While the
  presence of a SN is clear, the SN itself is not yet fully
  characterised.}  These SNe share the characteristics of SN~1998bw,
in the sense that they are of type Ic and they have broad lines,
indicative of high expansion velocity.

There were attempts to obtain a spectrum of the SN for a number of
other GRBs but the evidence is not as solid as one would wish.  The
reader is referred to \cite{HB12} for a detailed discussion of each
event and associated evidence.

\subsection{Light curves bumps}

Short of obtaining a spectrum around the time of the SN peak
luminosity, a well-sampled light curve in several filters can also
show evidence for a SN. This will take the form of a ``slowing `down''
of the light curve or even a rebrightening. A fairly large number of
bursts have displayed such a behaviour and in most cases, the bump can
be reasonably well fit with a SN light curve (see
Table~\ref{tab:sne}).  These SNe are represented as ``modified
SN~1998bw'', in the sense that the adopted model light curve is
usually that of SN~1998bw, stretched or compressed in time and shifted
in brightness (and properly accounting for the different luminosity
distance and \textit{k}-correction).

The brightness of these bumps is usually within a factor of a few of
SN~1998bw (e.g. \cite{Zeh04, Ferrero06, Cano11a}) albeit slightly
fainter on average, which is not surprising given that SN~1998bw was a
very bright event.
At this point, one can say that in nearly every case where we should
have been able to see a bump, a bump has been seen and it corresponds,
in brightness and timing, to a SN.

\begin{table}
\begin{tabular}{l}
\hline
\hline
\textbf{a)} GRBs with highest quality spectroscopic evidence:\\
GRB980425 $\equiv$ SN~1998bw, GRB030329 $\equiv$ SN~2003dh, GRB031203
$\equiv$ SN~2003lw, \\
GRB060218 $\equiv$ SN~2006aj, GRB100316D $\equiv$ SN~2010bh, GRB091127 
$\equiv$ SN~2009nz, \\
GRB120422A $\equiv$ SN~2012bz \\
\hline
\textbf{b)} Other GRBs with SN evidence: \\
GRB970228, GRB980326, GRB990712, GRB991208, GRB000911, GRB011121,\\
GRB020305, GRB020405, GRB020410, GRB020903, GRB021211, GRB030723,\\
GRB040624, GRB041006, GRB050416A, GRB050525A, GRB050824, GRB060729,\\
GRB070419A, GRB080319B, GRB081007, GRB090618, GRB100418A, GRB101219B,\\
GRB101225A\\
\hline
\hline
\end{tabular}
\caption{A list of GRBs where some evidence for a SN has been found
  (see Table~9.1 in \cite{HB12} for references related to any
  particular GRB). The first category includes those events with a
  very clear spectroscopic signature; the second category includes
  event where the evidence ranges from good to poor.}
\label{tab:sne}
\end{table}

\subsection{Properties of SNe of GRBs}

We want to understand the explosions so that we can ultimately
constrain the progenitors of GRBs. A good multi-colour light curve,
coupled with a velocity measurement at maximum light, can be modelled
analytically \cite{Arnett82} to obtain explosion parameters: kinetic
energy $E_k$, mass of the ejecta $M_{ej}$, and mass of $^{56}$Ni
synthesised. This has been done for a few events (see \cite{Cano11b,
  Olivares12, Bufano12} for the case of XRF100316D). This approach may
be subject to systematic errors however and modelling based on light
curve and time-series spectroscopy yields more robust results
(e.g. \cite{Iwamoto98, Mazzali03, Maeda07}).  This provides a better
understanding of the ejecta's structure and yields more reliable
results than simple light curve analysis because it avoids several
simplifications.
The results of such modelling give a range of values for each parameter:
the kinetic energy is $E_k = 2-60 \times 10^{51}$ erg, the ejected
mass ranges from 2 $M_\odot$ to 13 $M_\odot$ and the mass of $^{56}$Ni
synthesised in the explosion is between 0.1 $M_\odot$ and 0.7
$M_\odot$.

\section{Cautionary tales}

In some cases, the observational data may not allow us to detect a SN
signature.  One such example is GRB070419A. While there was an initial
claim for a SN \cite{Hill09} based on a photometric bump, a full
analysis of all the light curve data showed that the late time
behaviour could be fully explained by a fairly shallow decay of the
afterglow \cite{Melandri09}. In order to determine the SN brightness,
we need to know very well what the afterglow is doing. This means that
a well-sampled GRB light curve is necessary for tens of days; not only
for a few hours.  Observing the source past the SN maximum is needed
as well; if any ``rebrightening'' is seen, the fast fading of the SN
will leave a different signature in the light curve than the power
law decay of the afterglow.

\begin{figure}
\includegraphics[angle=270, scale=0.55]{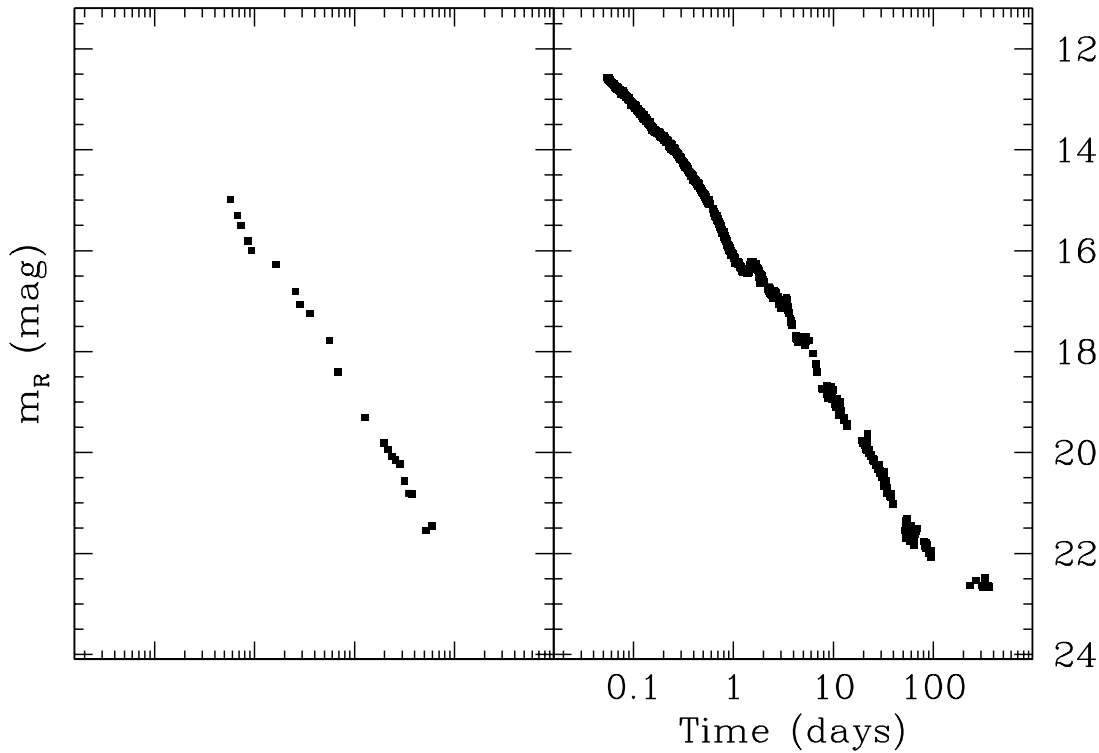}
\caption{\emph{Left:} An $R$-band light curve for a seemingly
  anonymous GRB. The afterglow decays by several magnitudes over $\sim
  2.5$ decades in time. At first sight, there is no SN bump.
  \emph{Right:} The whole $R$-band light curve of GRB030329 which is
  the event on the left. There is no doubt that there was a SN
  associated with this event.}
\label{fig:lc0329}
\end{figure}

Another example is given in Fig.~\ref{fig:lc0329}. The left panel is a
light curve displaying a more or less power law decay. From this light
curve, it is impossible to tell whether there is a SN hidden in the
data or not.
The right panel of the figure shows the full light curve, which is for
GRB030329. Now, obviously a SN was there to be found in the data. 
Assume for a moment that this burst had been at $z\sim 0.5$, hence
much fainter, we would have had much less data, the light curve would
have looked like the left panel of Fig.~\ref{fig:lc0329} and it would
have been very hard to find a SN in the light curve. The point here is
that even if we do not see an obvious SN bump, there may well be one
but we need good enough data to separate the relative contributions of
the afterglow and SN. The case of GRB060729 is particularly
instructive: there is no SN bump to be seen simply because the
afterglow was very bright but a careful analysis revealed a bright SN
in the light curve \cite{Cano11a}.

Another example is provided by GRB101225. There are two ``competing''
interpretations regarding the nature of this event.  One is a
cosmological GRB ($z\simeq 0.33$) with a SN bump \cite{Thoene11}, the
other is a minor planetary body tidally disrupted by a (Galactic)
neutron star \cite{Campana11}.
The point here is not to decide which interpretation is best but to
emphasise again that \emph{the right type of data at the right time}
will go a long way towards removing any ambiguities in the analysis.
This burst serves as a stark reminder that nothing replaces
spectroscopy.

\section{GRBs or SNe we do not see}

The central point of this section is about negatives: not finding what
we expect or finding what we do not expect.

\subsection{Observing a GRB and finding no SN}

There are times when, try as you may, there is no SN to be found. Two
events in particular, GRB060505 and GRB060614, are now etched in the
collective consciousness.  Both were at a redshift low enough that a
SN even substantially fainter than SN~1998bw would have been detected
easily. The limits on the brightness of any SN are actually very
strong: in both cases any SN would have to have been at least 100
times fainter than SN~1998bw at peak (e.g. \cite{Fynbo06, Gal-Yam06,
  DellaValle06}).  Several other events also showed no sign of a SN,
although in some cases the lack of redshift makes the evidence weaker
(\cite{Soderberg05, Levan05}).

While there may still be some lingering questions regarding the nature
of these bursts, they may be in a different class than most long GRBs.
A possibility is that they have slightly different
progenitors. Indeed, it may well be possible to make a GRB (i.e. a
relativistic jet) without making a bright SN (e.g. \cite{Woosley93,
  Tominaga07, Moriya10}). A bright burst of $\gamma$ rays may be the
only electromagnetic signature of such events.  Obviously we need to
find more of these ``failed'' SNe. This is why it is important to
monitor low redshift GRBs for several weeks. The absence of a SN may
tell us even more than an actual detection.

\subsection{Observing a SN and serendipitously finding a GRB}

Given that GRBs emit their radiation in a narrow cone, we miss most of
them (i.e. those pointing away from Earth). The emission becomes
visible however at late times in the radio when the afterglow is
emitting essentially isotropically.  Because we can separate normal
SNe from late-time jet emission via their luminosities (jets are much
more luminous than SNe, e.g. \cite{Soderberg10} and references
therein), this offers a way of finding GRBs whose high-energy emission
was directed away from us.

Observations of SNe at radio wavelengths have recently uncovered such
a case.  The radio luminosity of SN~2009bb places it squarely among
the jet-associated SNe \cite{Soderberg10}, yet no GRB was
observed. Another possible case is SN~2007gr \cite{Paragi10}.
These recent observations of nearby SNe show that the number of jets
found via their SN emission may be comparable to the number of SNe
found via their early gamma-ray emission.  If the two techniques have
comparable yields, they should both be pursued as each complements the
other.

\section{The SNe of GRBs in a general context}

From their relative rates, we know that GRBs represent a very small
fraction of all SNe. From their observed properties however, they do
stretch the parameter space of stellar explosions.  
We know that local ``normal'' SNe and GRBs mark the explosions of
massive stars. What sets GRBs apart from other SNe is the large amount
of energy that goes into relativistic ejecta.  What we ultimately want
to know is why a small fraction of SNe manage to do that while most of
them do not. Comparing the respective observational properties of these
two types of explosions is a fruitful way of understanding their
differences.

\begin{figure}
\begin{center}
  \includegraphics[width=.6\textwidth]{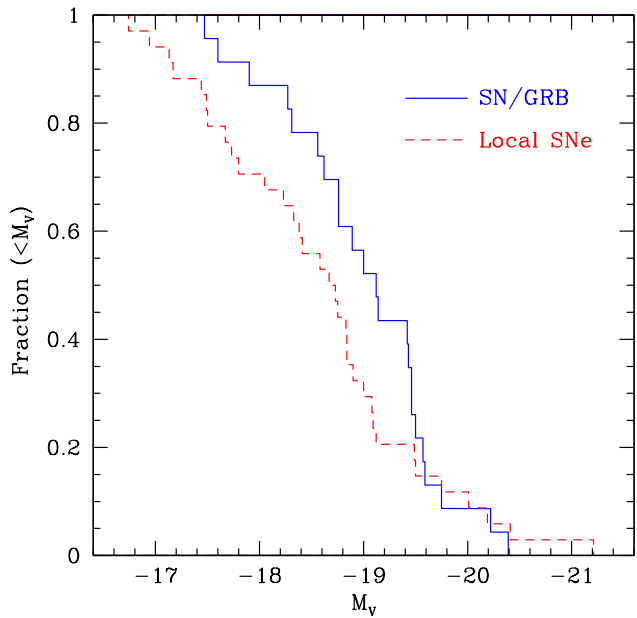}
\end{center}
\vspace*{-10mm}
\caption{Cumulative distribution of peak magnitudes for local SNe Ibc
  (red dashed line) and the SNe associated with GRBs (blue solid
  line). On average, the SNe of GRBs are brighter than local SNe Ibc
  (based on data in \cite{Cano11a}). This should be interpreted with
  some caution however as observations of a SN/GRB at an intermediate
  redshift (say $z\simeq 0.3-0.5$) is easier when the SN is bright. }
\label{fig:peak_mag}
\end{figure}

One similarity is the spectroscopic type, Ic, which tells us that GRB
progenitors are stripped-envelope stars.  They also share the same
``locations'' on the respective host galaxies (\cite{Fruchter06,
  Kelly08}): local SNe~Ic and GRBs are much more concentrated on the
light of their hosts than other types of SNe.  The SNe of GRBs show,
on average, larger ejection velocities than local SNe~Ic, they have
very broad lines. They also have larger energies and peak brightnesses
than local SNe~Ic (see Fig.~\ref{fig:peak_mag}).

The properties of GRB host galaxies are instructive as well; they are
different from those of SNe~Ic. GRBs are found in relatively low
luminosity irregular galaxies \cite{Fruchter06}. GRB are also found in
more metal-poor environments than SNe~Ic (e.g. \cite{Kelly11,
  Modjaz11}, see also \cite{Savaglio11} however). In this context, it
is interesting to note that local \emph{broad-line} SNe~Ic tend to
prefer low metallicity environments as well (\cite{Modjaz11}).

\section{Conclusion}

A bump on a light curve in itself is moderately interesting; it
becomes much more valuable when combined with time-series spectroscopy
because this allows us to characterise the progenitor star.
Photometric monitoring low-$z$ GRB is warranted however because this
is the only way to find other SN-less GRB (like GRB060614). It is
important to know what fraction of bursts can be formed via this
channel.  This will eventually help us understand whether these
objects are a separate class altogether or part of a continuum
encompassing very faint and very bright SNe.

It can be frustrating to wait for nature to be so kind as to offer us
a low redshift GRB where the SN can be studied in detail. Observing
nearby SNe can help us tremendously to understand GRBs in general.
Radio monitoring of SNe has been proved to be a good way to find
relativistic ejecta in the local universe. On the other hand, ongoing
optical surveys of the sky (PTF, Pan-STARRS, SkyMapper, La Silla
QUEST, CRTS, etc.) already find hundreds of SNe of all types every year.
This will lead to a better understanding of the evolution and
explosion of massive stars that will improve our understanding of
GRBs.
This is like making a jigsaw puzzle with a 1000 pieces with one blank
piece among them. Obviously we do not know where it fits in the global
picture but once the other 999 pieces are in place, we will will know
exactly where the blank one fits and, looking at the general picture,
we will have a very good idea as to what this piece should look like.


\begin{thebibliography}{99}

%
%
%
%
%
%


\bibitem{Arnett82} W.~D.~Arnett, 
\emph{Type I supernovae. I - Analytic solutions for the early part of
  the light curve}, 
\emph{ApJ} {\bf 253} (1982) 785

\bibitem{Berger11} E.~Berger, et al.,
\emph{The Spectroscopic Classification and Explosion Properties of SN
  2009nz Associated with GRB 091127 at z = 0.490},
\emph{ApJ} {\bf 743} (2011) 204

\bibitem{Bufano12} F.~Bufano,  et al.,
\emph{The Highly Energetic Expansion of SN2010bh Associated with GRB
  100316D}, 
\emph{ApJ} {\bf in press} (2012) [{\tt arXiv:1111.4527}]

\bibitem{Campana11} S.~Campana, et al.,
\emph{The unusual gamma-ray burst GRB 101225A explained as a minor
  body falling onto a neutron star}, 
\emph{Nature} {\bf 480} (2011) 69

\bibitem{Cano11a} Z.~Cano, et al., 
\emph{A tale of two GRB-SNe at a common redshift of z = 0.54},
\emph{MNRAS} {\bf 413} (2011) 669

\bibitem{Cano11b} Z.~Cano,  et al., 
\emph{XRF100316D/SN 2011bh and the nature of gamma-ray burst
  supernovae}, 
\emph{ApJ} {\bf 740} (2011) 41

\bibitem{Chornock10} R.~Chornock, et al., 
\emph{Spectroscopic Discovery of the Broad-Lined Type Ic Supernova
  2010bh Associated with the Low-Redshift GRB 100316D},
\emph{ApJ} {\bf submitted} (2010)  [{\tt arXiv:1004.2262}]

\bibitem{Colgate68} S.~A.~Colgate,
\emph{Prompt gamma rays and X-rays from supernovae},
\emph{CaJPh} {\bf 46} (1968) 476

\bibitem{DellaValle06} M.~Della Valle, et al., 
\emph{An enigmatic long-lasting $\gamma$-ray burst not accompanied by a
  bright supernova},
\emph{Nature} {\bf 444} (2006) 1050

\bibitem{Ferrero06} P.~Ferrero, et al. 
\emph{The GRB 060218/SN 2006aj event in the context of other gamma-ray
  burst supernovae},
\emph{A\&A} {\bf 457} (2006) 857

\bibitem{Fruchter06} A.~S.~Fruchter, et al. 
\emph{Long $\gamma$-ray bursts and core-collapse supernovae have
  different environments}, 
\emph{Nature} {\bf 441} (2006) 463

\bibitem{Galama98} T.~J.~Galama, et al. 
\emph{An unusual supernova in the error box of the $\gamma$-ray burst of 25
  April 1998},
\emph{Nature} {\bf 395} (1998) 670

\bibitem{Fynbo06} J.~P.~U.~Fynbo, et al.,
\emph{No supernovae associated with two long-duration $\gamma$-ray bursts},
\emph{Nature} {\bf 444} (2006)  1047

\bibitem{Gal-Yam06} A.~Gal-Yam, et al., 
\emph{A novel explosive process is required for the γ-ray burst GRB
  060614},
\emph{Nature} {\bf 444} (2006) 1053

\bibitem{Hill09} J.~Hill, et al., 
\emph{GRB 070419A, deep LBT photometry and possible supernova detection},
\emph{GCN Circulars} {\bf 6486} (2007)

\bibitem{Hjorth03} J.~Hjorth, et al.,
\emph{A very energetic supernova associated with the $\gamma$-ray
  burst of 29 March 2003},
\emph{Nature} {\bf 423} (2003) 847

\bibitem{HB12} J.~Hjorth, J.~S.~Bloom, \emph{Chapter 9 of ``Gamma-Ray
  Bursts'' eds. C.~Kouveliotou, S.~E.~Woosley, R.~A.~M.~J.~Wijers},
  Cambridge University Press, Cambridge, 2012 [{\tt
      astro-ph/1104.2274}]

\bibitem{Iwamoto98} K.~Iwamoto, et al., 
\emph{A hypernova model for the supernova associated with the $\gamma$-ray
  burst of 25 April 1998}, 
\emph{Nature,} {\bf 395} (1998) 672

\bibitem{Kelly08} P.~L.~Kelly, R.~P.~Kirshner, M.~Pahre, 
\emph{Long $\gamma$-Ray Bursts and Type Ic Core-Collapse Supernovae
  Have Similar Locations in Hosts},
\emph{ApJ} {\bf 687} (2008) 1201

\bibitem{Kelly11} P.~L.~Kelly, R.~P.~Kirshner,
\emph{Core-Collapse Supernovae and Host Galaxy Stellar Populations},
\emph{ApJ submitted} (2012) [{\tt arXiv:1110.1377}]

\bibitem{KSO73} R.~W.~Klebesadel, I.~B.~Strong, R.~A.~Olson,
\emph{Observations of Gamma-Ray Bursts of Cosmic Origin},
\emph{ApJ} {\bf 182} (1973) L85

\bibitem{Levan05} A.~J.~Levan, et al., 
\emph{A Deep Search with the Hubble Space Telescope for Late-Time
  Supernova Signatures in the Hosts of XRF 011030 and XRF 020427}
\emph{ApJ} {\bf 622} (2005) 977

\bibitem{Maeda07} K.~Maeda, et al., 
\emph{SN 2006aj Associated with XRF 060218 at Late Phases:
  Nucleosynthesis Signature of a Neutron Star-driven Explosion},
\emph{ApJ} {\bf 658} (2007) L5


\bibitem{Malesani04} D.~Malesani, et al., 
\emph{SN 2003lw and GRB 031203: A Bright Supernova for a Faint
  Gamma-Ray Burst},
\emph{ApJ} {\bf 609} (2004) L5

\bibitem{Malesani12} D.~Malesani, et al., 
\emph{GRB 120422A: VLT/X-shooter spectroscopic evidence for a SN},
\emph{GCN Circulars} {\bf 13277} (2012)

\bibitem{Matheson03} T.~Matheson,  et al., 
\emph{Photometry and Spectroscopy of GRB 030329 and Its Associated
  Supernova 2003dh: The First Two Months},
\emph{ApJ} {\bf 599} (2003) 394

\bibitem{Mazzali03} P.~A.~Mazzali, et al.,
\emph{The Type Ic Hypernova SN 2003dh/GRB 030329},
\emph{ApJ} {\bf 599} (2003) L95

\bibitem{Melandri09} A.~Melandri, et al., 
\emph{Evidence for energy injection and a fine-tuned central engine at
  optical wavelengths in GRB 070419A},
\emph{MNRAS} {\bf 395} (2009) 1941

\bibitem{Modjaz11} M.~Modjaz, et al., 
\emph{Progenitor Diagnostics for Stripped Core-collapse Supernovae:
  Measured Metallicities at Explosion Sites},
\emph{ApJ} {\bf 731} (2011) L4

\bibitem{Moriya10} T.~Moriya,  et al., 
\emph{Fallback supernovae: A possible origin of peculiar supernovae
  with extremely low explosion energies}
\emph{ApJ}  {\bf 719} (2010) 1445

\bibitem{Olivares12} F.~Olivares~E., et al.,
\emph{The fast evolution of SN~2010bh associated with XRF100316D},
\emph{A\&A} {\bf 539} (2012) A76

\bibitem{Paragi10} Z.~Paragi, et al., 
\emph{A mildly relativistic radio jet from the otherwise normal type
  Ic supernovae 2007gr}, 
\emph{Nature} {\bf 463} (2010) 516


\bibitem{Pian06} E.~Pian, et al., 
\emph{An optical supernova associated with the X-ray flash XRF060218},
\emph{Nature} {\bf 442} (2006) 1011

\bibitem{Sanchez12} R.~Sanchez-Ramirez, et al.,
\emph{GRB~120422A: SN identification from GTC},
\emph{GCN Circulars} {\bf 13281} (2012)

\bibitem{Savaglio11} S.~Savaglio, et al.,
\emph{Supersolar metal abundances in two galaxies at z$\sim$3.57
  revealed by the GRB090323 afterglow spectrum}, 
\emph{MNRAS} {\bf 420} (2011) 627

\bibitem{Soderberg05} A.~M.~Soderberg, et al., 
\emph{An HST Search for Supernovae Accompanying X-Ray Flashes}
\emph{ApJ} {\bf 627} (2005) 877

\bibitem{Soderberg10} A.~M.~Soderberg,  et al., 
\emph{A relativistic type Ibc supernova without a detected $\gamma$-ray
  burst},
\emph{Nature} {\bf 463} (2010) 513

\bibitem{Stanek03} K.~Z.~Stanek, et al.,
\emph{Spectroscopic discovery of the supernova 2003dh associated with
  GRB 030329},
\emph{ApJ} {\bf 591} (2003) L17

\bibitem{Thoene11} C.~C.~Th\"one, et al.,
\emph{The unusual $\gamma-$ray burst GRB 101225A from a helium
  star/neutron star merger at redshift 0.33},
\emph{Nature} {\bf 480} (2011) 72

\bibitem{Tominaga07} N.~Tominaga,  et al., 
\emph{The Connection between Gamma-Ray Bursts and Extremely Metal-poor
  Stars: Black Hole–forming Supernovae with Relativistic Jets},
\emph{ApJ} {\bf 657} (2007) L77

\bibitem{Woosley93} S.~E.~Woosley,
\emph{Gamma-ray bursts from stellar mass accretion disks around black holes}
\emph{ApJ} {\bf 405} (1993) 273 

\bibitem{WB06} S.~E.~Woosley, J.~S.~Bloom,
\emph{The Supernova Gamma-Ray Burst Connection},
\emph{ARA\&A}{\bf  44} (2006) 507

\bibitem{Zeh04} A.~Zeh, S.~Klose, D.~H.~Hartmann,
\emph{A Systematic Analysis of Supernova Light in Gamma-Ray Burst
  Afterglows},
\emph{ApJ} {\bf 609} (2004) 952




\end{thebibliography}
\end{document}